\documentstyle[11pt]{article}
\begin{document}

\begin{centering}
{\Large High resolution spectroscopy of muonium}\\
\end{centering}
\vspace*{5mm}

\begin{centering}
{K.~Jungmann $^*$\\
\vspace*{3mm}
Physikalisches Institut der Universit\"at
Heidelberg\\
Philosophenweg 12\\
 D-69120 Heidelberg, Germany\\}
\end{centering}

\vspace*{10mm}

The hydrogen-like muonium atom ($\mu^+e^-$) consists of a positive muon
($\mu^+$) and an electron ($e^-$).
Since it was first observed by Hughes et al. in 1960 [1], a series of precision
experiments could be carried out testing bound state Quantum Electrodynamics
(QED) and the nature of the muon as a point-like leptonic particle [2].
Precision values for fundamental constants could be obtained,
including the fine structure constant $\alpha$ and the muon's magnetic
moment $\mu_{\mu}$ [2].
Since no internal structure could be revealed for leptons down to dimensions
of $10^{-18}$m,
the accuracy of theoretical calculations of term energies is higher
than for the natural hydrogen atom and its isotopes deuterium and tritium
as well as exotic hydrogen-like systems which contain hadrons.
Internal structure effects of these particles
spoil the precision of theoretical results.
Of particular interest for precision muonium experiments, which are carried
out in order to test standard
theory and to extract accurate
values for fundamental constants,
are the ground state hyperfine structure splitting $\Delta \nu _{HFS}$ [3]
and the 1s-2s interval  $\Delta \nu _{1s2s}$ [4].
Both cases involve the n=1 ground state
in which the atoms can be produced in sufficient quantities.\\

In the case of the ground state hyperfine 
structure splitting the present experimental accuracy
[3]
is about half the size of the strong interaction contributions and about
three times as large as the contribution from weak interaction.
At the Los Alamos Meson Physics Facility, LAMPF (USA), an
experiment is  performed by a Yale-Heidelberg-LAMPF-Sussex collaboration
which aims for an improved determination of  $\Delta \nu _{HFS}$
and the muon's magnetic moment $\mu_{\mu}$  from microwave measurements
of two Zeeman transitions in a 1.7 T magnetic field [5].
The experiment uses the "old muonium" technique to obtain
resonance lines narrower than the natural linewidth.
At LAMPF the surface muon beam can be
chopped in bunches which are comparable in their length to
the muon's lifetime of $\tau_{\mu}=2.20 \mu s$, typically $4 \mu s$.
A substantially reduced linewidth of the transitions can be obtained,
if the signal is observed only after a period which is long compared
to $\tau_{\mu}$. An all the time fixed phase relationship between
the atoms and the microwave field, to which the atoms are exposed,
is essential. Line widths (FWHM) 
as narrow as half the natural line width of 145 kHz
could be recorded [5]. The experiment was carried out at LAMPF both with 
magnetic field sweep and with microwave frequency scanning. The analysis
is close to be completed and about half an order of magnitude improvement
is expected for both for the zero field ground state hyperfine splitting and
the magnetic moment of the muon [6]. An improved value for the mass of 
the muon can be expected. The experiment appears to be mainly limited 
by statistics. \\

In the muonium atom the optical 1S-2S two-photon transition offers the
highest resolution of all possible
electromagnetic transitions.
An experiment to precisely measure
the 1s-2s frequency interval has been started
 at the worldwide brightest pulsed surface muon
 source at present, the ISIS facility of the
 Rutherford Appleton Laboratory (RAL) in Chilton, UK,
by a Heidelberg-Oxford-Rutherford-Sussex-Yale-Novosibirsk
collaboration [4,7].
The muonium atoms are produced by stopping a muon beam
 close to the surface of a SiO$_2$ powder target.
 Every 20~msec 85(15) muonium atoms in the 1S state leave the target surface
 with thermal velocities into the adjacent vacuum region.
 The F=1 component of the $1^2S_{1/2}-2^2S_{1/2}$
 transition could be excited
 by Doppler-free two-photon laser spectroscopy with a pulsed laser
beam generated by frequency doubling of the
output of an excimer laser pumped dye laser amplifier
which was fed from a cw ring dye laser.
The two-photon transition  was observed by
 the
 photoionization of the 2S state
 with a third photon from the same laser field and the subsequent
detection of the slow muon released in the process.
The presently reached experimental accuracy is approximately the size
of the difference in the Lamb shift contributions between hydrogen and
muonium. The experimental uncertainty is dominated by properties
of the necessary intense UV laser light. First there is an ac Stark shift
and second a frequency chirping arose during the fast pumping process
of the laser amplifier.
A new laser laser system consisting of a Ti::sapphire laser seeded
alexandrite ring laser amplifier, the output of which will be frequency tripled,
promises more than an order of magnitude lower chirp and higher
experimental accuracy. It is an essential feature of the laser system that
time dependent phse fluctuations, which have been the source of the major 
contribution to the systematic errors in previous measurements, are 
compensated at the MHz level 
by means of an electro-optic device in the resonator of the 
ring laser amplifier. 
The experiment aims has had a successful run in December 1997 in which 
at resonance a signal rate of 9 events per hour was observed for 3 MJ laser 
energy at 244~nm wavelength. The experiment ultimately aims 
for a determination
of the 1s-2s interval at the  MHz level. At this accuracy the most precise 
value of the muon mass can be extracted. \\

An improved knowledge of the muon mass is essential for the success
of a new measurement of the muon's magnetic anomaly, which may contain
contributions from physics beyond the standard model and promises a clean
test of the renormalizability of electroweak interactions [8].
Such an experiment is being set up at the Brookhaven National Laboratory,
Upton, USA.\\

The accuracy of both spectroscopic experiments on muonium 
described above could be
further improved at a pulsed surface muon beam of significantly higher 
particle flux. Such beam could be provided by the JHF.
The microwave experiment would benefit 
from a pulsed muon source of pulses with up to $\approx$~4~$\mu$sec and
any pulse separation above 15~$\mu$sec 
The laser experiments, however, would  need a similar pulse length but
a repetition rate which matches the laser repetition rate which typically 
is of order 50~Hz. In addition to the muonium experiments, investigations of
fundamental atomic systems like muonic hydrogen and muonic helium
could be performed at levels of accuracy which could yield
important information on properties of the respective nuclei
at low momentum transfer. A variety of possible experiments
in muonic atomic systems has been considered by Boshier et al.
recently [8].\\

This work is supported in part 
by the German Bundesminister f\"ur Bildung und Forschung
(BMBF), the US Department of Energy (DOE), the British Engineering 
and Research Council
(EPSERC) and a NATO research grant.\\

\small
 $^*$ This work contains results achieved by the muonium hyperfine 
structure collaboration and the Muonium 1s-2s collaboration.

The muonium hyperfine structure collaboration at 
LAMPF consists of:\\
S. Dhawan,
V.W. Hughes,
D. Kawall,
W. Liu
M. Grosse-Perdekamp
{\it Yale University};
K.~Jungmann,
G.~zu~Putlitz,
I.~Reinhard,
{\it University of Heidelberg};
M.G. Boshier
{\it University of Sussex};
O. van Dyck,
C. Pillai,
{\it LAMPF}
F.G. Mariam,
{\it Brookhaven National Laboratory};
 \newline
The muonium 1s-2s collaboration at RAL consists of:\\
A. Gro\ss{}mann,
K.~Jungmann,
J. Merkel,
V. Meyer,
G.~zu~Putlitz,
I.~Reinhard,
R.~Santra,
K. Tr\"ager,
L. Willmann,
{\it University of Heidelberg};
P.E.G. Baird,
P. Bakule,
S. Cornish,
I. Lane,
L.W. Lin,
P.G.H. Sandars,
{\it University of Oxford}
W.T. Toner,
G.H. Eaton,
C.A. Scott,
M. Towrie,
{\it Rutherford Appleton Laboratory};
M.G. Boshier
{\it University of Sussex};
V.W. Hughes,
{\it Yale University};
S.N. Bagayev,
A.~Dyschkhoff
Y. Matyugin,
{\it ILP Novosibirsk}
\vspace*{10mm}

{\normalsize \bf References}
\vspace*{5mm}
\noindent
\begin{itemize}
 \item[1]
V.W. Hughes, D.W. McColm, K. Ziock, and R. Prepost, Phys.Rev.Lett.{\bf5}
(1960) 63.
 \item[2] V.W.~Hughes and G.~zu~Putlitz, in: "Quantum
Electrodynamics", p. 822ff, T.~Kinoshita (ed.), World
Scientific, Singapore
(1990) and references therein.
 \item[3]
F.G. Mariam, W. Beer, P.R. Bolton, P.O. Egan, C. J.
Gardner, V.W. Hughes, D.C. Lu, P.A. Souder,
H.~Orth, J. Vetter, U.
Moser, and G.~zu~Putlitz, Phys. Rev. Lett. {\bf 49} (1982) 993.
 \item[4]
F. Maas, P.E.G. Baird, J.R.M. Barr, D. Berekeland, M.G. Boshier, B. Braun,
G.H. Eaton, A.I.
Ferguson, H. Geerds, V.W. Hughes, K. Jungmann,
B.M. Matthias, P. Matousek,
M.A. Persaud, G.~zu~Putlitz, I. Reinhard, E. Riis,
P.G.H. Sandars, W. Schwarz,
W.T. Toner, M. Towrie, L.
Willmann, K.A. Woodle,
G.Woodman, and L. Zhang,
Phys.Lett.A{\bf187} (1994) 247;
see also: S. Chu, A.P. Mills,
A.G. Yodh, K. Nagamine, Y. Miyake,
T. Kuga, Phys. Rev. Lett. {60}, 101 (1988)
 \item[5]
M.G. Boshier, S. Dhawan, X. Fei, V.W. Hughes, M. Janousch, K. Jungmann,
W. Liu, C. Pillai, R.
 Prigl, G. zu Putlitz, I. Reinhard, W. Schwarz,
P.A. Souder, O. van Dyck, X. Wang, K.A. Woodle, Q.
Xu,
Phys.Rev. {\bf A52} (1995) 1948.
 \item[6] V.W. Hughes, in: Atomic Physics Methods in Modern Research, 
K. Jungmann, J. Kowalski, I. Reinhard and F. Tr\"ager (eds.), Springer, 
Heidelberg (1997) p. 21.
 \item[7]
K.~Jungmann, in:
Atomic Physics
14, D.J. Wineland, C.E. Wieman and S.J. Smith (eds.),
AIP Press,
New York, p. 102 (1994) and Phys.Bl. {\bf 51}, 1159 (1995).
 \item[8]
B.L.~Roberts, Z.Phys.C {\bf56} (1992) S109 , and
V.W.~Hughes, in: Gift of Prophecy, E.C.G.~Sudarshan
(ed.), World Scientific, Singapore, (1995) p. 222.
 \item[9]
M.G.~Boshier, Comm. At.Mol.Phys. {\bf 33}, 17 (1996).
\end{itemize}

\end{document}